\documentclass[manuscript]{aastex}
\usepackage[export]{adjustbox}
\usepackage{ragged2e}

\usepackage{lineno}

\usepackage{xcolor}

\shorttitle{Evolution of an helical structure  observed by Metis during the eruption of a polar crown prominence}
\shortauthors{Romano et al.}

\begin{document}
\justifying

\title{
Metis Observations of Alfv\'{e}nic Outflows Driven by Interchange Reconnection in a Pseudostreamer
}

\author{P. Romano\altaffilmark{1}, 
P. Wyper\altaffilmark{2}, 
V. Andretta\altaffilmark{3},
S. Antiochos\altaffilmark{4},
G. Russano\altaffilmark{3}, 
D. Spadaro\altaffilmark{1},  
L. Abbo\altaffilmark{5}, 
L. Contarino\altaffilmark{1},
A. Elmhamdi\altaffilmark{7},
F. Ferrente\altaffilmark{1},
R. Lionello\altaffilmark{8}, 
B.J. Lynch\altaffilmark{9}, 
P. MacNeice\altaffilmark{6}, 
M. Romoli\altaffilmark{10,11},
R. Ventura\altaffilmark{1}, 
N. Viall\altaffilmark{6}, 
A. Bemporad\altaffilmark{5},
A. Burtovoi\altaffilmark{10,5},
V. Da Deppo\altaffilmark{12},
Y. De Leo\altaffilmark{1,13},
S. Fineschi\altaffilmark{5},
F. Frassati\altaffilmark{5},
S. Giordano\altaffilmark{5},
S.L. Guglielmino\altaffilmark{1},
C. Grimani\altaffilmark{14,15}, 
P. Heinzel\altaffilmark{16,17},
G. Jerse\altaffilmark{18}, 
F. Landini\altaffilmark{5},
G. Naletto\altaffilmark{19}, 
M. Pancrazzi\altaffilmark{5}, 
C. Sasso\altaffilmark{3},
M. Stangalini\altaffilmark{20},
R. Susino\altaffilmark{5},
D. Telloni\altaffilmark{5},
L. Teriaca\altaffilmark{21}, 
M. Uslenghi\altaffilmark{22}}


\email{paolo.romano@inaf.it}



\altaffiltext{1}{National Institute for Astrophysics, Astrophysical Observatory of Catania, Via Santa Sofia 78, I-95123 Catania, Italy}
\altaffiltext{2}{Department of Mathematical Sciences, Durham University, Durham, DH1 3LE, UK}
\altaffiltext{3}{National Institute for Astrophysics, Astronomical Observatory of Capodimonte, Salita Moiariello 16, I-80131 Napoli, Italy}
\altaffiltext{4}{CLaSP, University of Michigan, Ann Arbor, MI 48109, USA}
\altaffiltext{5}{National Institute for Astrophysics, Astrophysical Observatory of Torino, Via Osservatorio 20, I-10025 Pino Torinese, Italy}
\altaffiltext{6}{Heliophysics Science Division, NASA Goddard Space Flight Center, 8800 Greenbelt Road, Greenbelt, MD 20771, USA}
\altaffiltext{7}{Department of Physics and Astronomy, King Saud University, PO Box 2455, Riyadh 11451, Saudi Arabia.}
\altaffiltext{8}{Predictive Science Inc., 9990 Mesa Rim Road, Suite 170, San Diego, CA 92121, USA}
\altaffiltext{9}{Department of Earth, Planetary, and Space Sciences, University of California–Los Angeles, Los Angeles, CA 90056, USA}
\altaffiltext{10}{University of Florence, Department of Physics and Astronomy, Via Giovanni Sansone 1, I-50019 Sesto Fiorentino, Italy}
\altaffiltext{11}{National Institute for Astrophysics, Astrophysical Observatory of Arcetri, Largo Enrico Fermi 5, I-50125 Firenze, Italy}
\altaffiltext{12}{National Research Council, Institute for Photonics and Nanotechnologies, Via Trasea 7, I-35131 Padova, Italy}
\altaffiltext{13}{Institute of Physics, University of Graz, Universitätsplatz 5, 8010, Graz, Austria}
\altaffiltext{14}{University of Urbino Carlo Bo, Department of Pure and Applied Sciences, Via Santa Chiara 27, I-61029 Urbino, Italy}
\altaffiltext{15}{National Institute for Nuclear Physics, Section in Florence, Via Bruno Rossi 1, I-50019 Sesto Fiorentino, Italy}
\altaffiltext{16}{Czech Academy of Sciences, Astronomical Institute, Fričova 298, CZ-25165 Ondrejov,Ondrejov, Czechia}
\altaffiltext{17}{University of Wrocław, Center of Scientific Excellence, Solar and Stellar Activity, ul. Kopernika 11, PL-51-622 Wrocaw, Poland}
\altaffiltext{18}{National Institute for Astrophysics, Astronomical Observatory of Trieste, Localit\'{a} Basovizza 302, I-34149 Trieste, Italy}
\altaffiltext{19}{University of Padua, Department of Physics and Astronomy, Via Francesco Marzolo 8, I-35131 Padova, Italy}
\altaffiltext{20}{Italian Space Agency, Via del Politecnico snc, I-00133 Roma, Italy}
\altaffiltext{21}{Max Planck Institute for Solar System Research, Justus-von-Liebig-Weg 3, D-37077 Göttingen, Germany}
\altaffiltext{22}{National Institute for Astrophysics, Institute of Space Astrophysics and Cosmic Physics of Milan, Via Alfonso Corti 12, I-20133 Milano, Italy} 
		 			


\begin{abstract}
This study presents observations of a large pseudostreamer solar eruption and, in particular, the post-eruption relaxation phase, as captured by Metis onboard the Solar Orbiter on October 12, 2022, during its perihelion passage. Utilizing total brightness data, we observe the outward propagation of helical features up to 3 solar radii along a radial column that appears to correspond to the stalk of the pseudostreamer. The helical structures persisted for more than 3 hours following a jet-like coronal mass ejection associated with a polar crown prominence eruption. A notable trend is revealed: the inclination of these features decreases as their polar angle and height increase. Additionally, we measured their helix pitch. Despite a 2-minute time cadence limiting direct correspondence among filamentary structures in consecutive frames, we find that the Metis helical structure may be interpreted as a consequence of twist (nonlinear torsional Alfv\'{e}n waves) and plasma liberated by interchange reconnection.  
A comparison was performed of the helix parameters as outlined by fine-scale outflow features with those obtained from synthetic white-light images derived from the high-resolution magnetohydrodynamics simulation of interchange reconnection in a pseudostreamer topology by \citet{Wyp22}. A remarkable similarity between the simulation-derived images and the observations was found.  We conjecture that these Metis observations may represent the upper end in spatial and energy scale of the interchange reconnection process that has been proposed recently as the origin of the Alfv\'{e}nic solar wind. 

\end{abstract}




\section{Introduction}
The Metis coronagraph \citep{Ant20, Fin20}, aboard the ESA's Solar Orbiter mission \citep{Mul20}, represents a major advancement in our ability to observe and understand the solar corona. Launched in February 2020, Solar Orbiter aims to study the Sun up close, and Metis plays a crucial role in this mission by providing high-resolution images of the outer corona in both visible and ultraviolet light. By occulting the bright solar disk, Metis enables scientists to observe the faint corona, unveiling intricate details of the Sun's outer atmosphere and its dynamic behavior \citep{Rom21}. The Metis coronagraph's advanced imaging capabilities and its integration with the Solar Orbiter's suite of instruments are enhancing our understanding of the coupling between the corona and solar wind, offering valuable data that supports both theoretical models and practical space weather applications.

Metis measures the brightness of the corona in two wavelength bands: broad-band visible light in the range 580 -- 640 nm (VL) and narrow-band ultraviolet radiation in the range 121.6$\pm$10~nm. The instrument is capable of observing at cadences that can be as high as 1 image per second, thus allowing the study of the structure and dynamics of the coronal plasma and magnetic fields, including the evolution of coronal mass ejections (CMEs), waves, and other transient events. 

Two important developments in recent years in the observation and theory of corona-solar wind coupling have made the Metis capabilities especially valuable. First, there is a growing consensus that the so-called Alfv\'{e}nic wind from coronal holes is due to ubiquitous jetting activity that is observed at the base of the corona. The jets are presumed to result from the interchange reconnection between the closed flux of small parasitic polarity regions and surrounding coronal hole open field \citep{Raouafi2023, Bale2023}.  Supporting evidence for this hypothesis comes from in situ measurements by the Parker Solar Probe (PSP), which demonstrate that the bulk of the solar wind exhibits fluctuations that are Alfv\'{e}nic in that the magnetic and velocity perturbations are tightly correlated \citep{Thepthong2024}. These fluctuations are believed to originate in the low corona and are a natural consequence of the interchange reconnection responsible for the jets. It should be noted that the idea of a reconnection driven origin for the wind is not new. \citet{Parker1992} and \citet{Axford1992} independently proposed that the heating and acceleration of the wind is due to reconnection between small-scale closed field regions and background open flux. The recent high-resolution observations of the corona and photosphere strongly support that hypothesis \citep{Raouafi2023, Bale2023, Chitta2024}. The ubiquitous jets drive both mass motions and Alfv\'{e}n waves onto open field lines, but due to their small scale only the mass flow has been imaged directly. We emphasize that these ubiquitous jets are well below the Metis field-of-view; consequently, Metis cannot test this model directly. Metis, however, can look for evidence of outflows from large-scale interchange reconnection events.

The second important development is high-resolution observations have also revealed that coronal jets are due to the eruption of a filament, just like large-scale eruptions such as CMEs \citep{Sterling2015}. It appears that in all cases the free energy to power eruptive events builds up at the base of the low corona in the form of the highly-sheared magnetic field of a filament channel overlying a polarity inversion line. In the case of coronal hole jets and pseudostreamer eruptions, the pre-eruption magnetic topology is that of a closed-field embedded bipole surrounded by open flux \citep[e.g.][]{Pariat2009}. This topology is physically identical to that of the breakout model for CME onset \citep{Antiochos1999}, in that it contains a dome-like separatrix surface in the corona with a null point. Reconnection at this coronal null point and separatrix allows the filament channel field to expand upward, causing a current sheet to form below the rising filament, which leads to strong flare reconnection there and eventually results in a jet-like eruption through the null points. We have argued and demonstrated with numerical simulations that this basic scenario is a universal mechanism for solar eruptions \citep{Wyper2017}. 

The key conclusion from these observational and theoretical developments is that we can probe the basic process that has been proposed for the origin of the Alfv\'{e}nic solar wind by studying pseudostreamer eruptions, which are large-scale and can be well-observed by Metis and other instruments.  This is the primary motivation for the work presented below. Our results capture these processes of interchange-driven plasma and Alfv\'{e}n wave injection from the corona to the heliosphere.   

The discussion above emphasizes the central role of erupting filament channels in powering the interchange reconnection that drives the wind. Although the pre-eruptive magnetic structure of filament channels is still under debate, there is widespread consensus that the final erupting structure that escapes from the corona as a CME, or that reconnects through a {\bf null point} as a jet, is a twisted flux rope due to the action of flare reconnection \citep{Pri02}. Note that by flux rope we refer to a magnetic flux tube in which the field lines spiral around a central axis.  Flux ropes are the fundamental structures that appear high in the corona during eruptive events and in situ measurements of interplanetary CMEs generally show a flux rope structue with significant twist/helicity \citep[e.g.][]{Dasso2005}. Metis observations have already proven effective for revealing the presence of eruptive flux ropes in the corona, providing evidence of their footpoints and helping to map their three-dimensional structure.  Flux ropes are known to play a critical role in the dynamics of the solar atmosphere \citep{Gib18} and the acceleration of CMEs \citep{Web12} and are often observed as filamentary structures in the solar corona \citep{Liu20}. 

In the context of CMEs, flux ropes are crucial because they carry away the magnetic free energy and helicity. The flux rope usually forms the bright core of a CME and in the standard MHKSP model for eruptive flares \citep{Car64, Stu66, Hir74, Kop76}, it is located above the flare current sheet in the traversal direction to the underlying cusp structure and is created by the flare reconnection \citep{Pri02}. When a flux rope erupts from inside a pseudostreamer, however, it generally undergoes interchange reconnection through the null with the surrounding open flux, so the resulting ejection is a jet or a narrow fan-like CME \citep{Wang2018,Kumar2021,Wyper2021}. The free energy and helicity contained in the flux rope is converted to mass acceleration and an Alfv\'{e}n wave flux by interchange reconnection, as required by the recent solar wind models \citep{Raouafi2023, Bale2023}.

Recent observations and simulations have highlighted the presence of helical and twisting structures in the solar corona, contributing to a better understanding of coronal dynamics and their relation to the solar wind. In particular, coronal jets and jetlets have been extensively studied as potential drivers of Alfv\'{e}nic disturbances propagating into the heliosphere. For instance, \citet{Ste20} discuss the possible evolution of minifilament-eruption driving coronal jets into magnetic twist-wave “switchbacks” observed by the PSP. Their study emphasizes that twisting motions observed in jet spires may propagate outward, forming magnetic structures detectable as transient magnetic field reversals in situ.

Furthermore, evidence from observational campaigns, including total and polarized brightness images, supports the interpretation that helical structures are a natural consequence of the reconnection process in the corona. For example, \citet{Hab14} provide insights into the polarimetric signatures of helical structures and their evolution in the corona, revealing key characteristics of these configurations at various altitudes.

In addition, white-light images from coronagraphs have shown the presence of propagating helical features consistent with theoretical predictions of magnetic flux rope dynamics in coronal mass ejections and jets. For example, \citet{Dru14} demonstrated the use of advanced image processing techniques to uncover fine-scale helical patterns in white-light observations, revealing the complexity of the coronal magnetic field's interaction with the solar wind. Similarly, \citet{Harrison2001} reported observations of twisted structures in the extended corona, interpreted as large-scale helical flux ropes forming in the aftermath of solar eruptions. These observations highlight that such helical configurations are not limited to small-scale jet-like structures but can also be found in larger-scale coronal phenomena, further supporting the role of magnetic reconnection in shaping the solar wind's properties.

In this work, we analyze high-cadence Metis observations taken in both unpolarized and polarized visible light, contextualized by EUI imaging, providing new insights into the nature of the corona - solar wind coupling \citep[see][]{And21, Bem22, Rus24}. These observations captured a jet-like CME on October 12, 2022, associated with the eruption of a polar crown prominence from beneath a pseudostreamer. We interpret the event in the framework of the breakout jet model \citep[e.g.][]{Wyper2018,Kumar2021,Wyper2021} and investigate the post-eruption relaxation phase, which exhibits evolving filamentary structures. To further understand these features, we compare Metis observations with synthetic white-light images derived from the high-resolution Magnetohydrodynamic simulation of slowly-driven interchange reconnection performed by \citet{Wyp22}, revealing similarities that offer new insights into coronal dynamics and the processes believed to drive the solar wind.

The Paper is organized as follows: the next section describes the data used and gives and overview of the observations. In section 3 we give our theoretical interpretation of the outflows observed by Metis. Section 4 details the methods employed to measure the inclination and pitch of the helical structures and the obtained results, section 5 outlines the comparison with the MHD simulation, and the final section presents the conclusions.


\section{The event of October 12, 2022 observed by Metis}

\begin{figure}
\begin{center}
\includegraphics[trim=0 0 0 40, clip, scale=0.64]{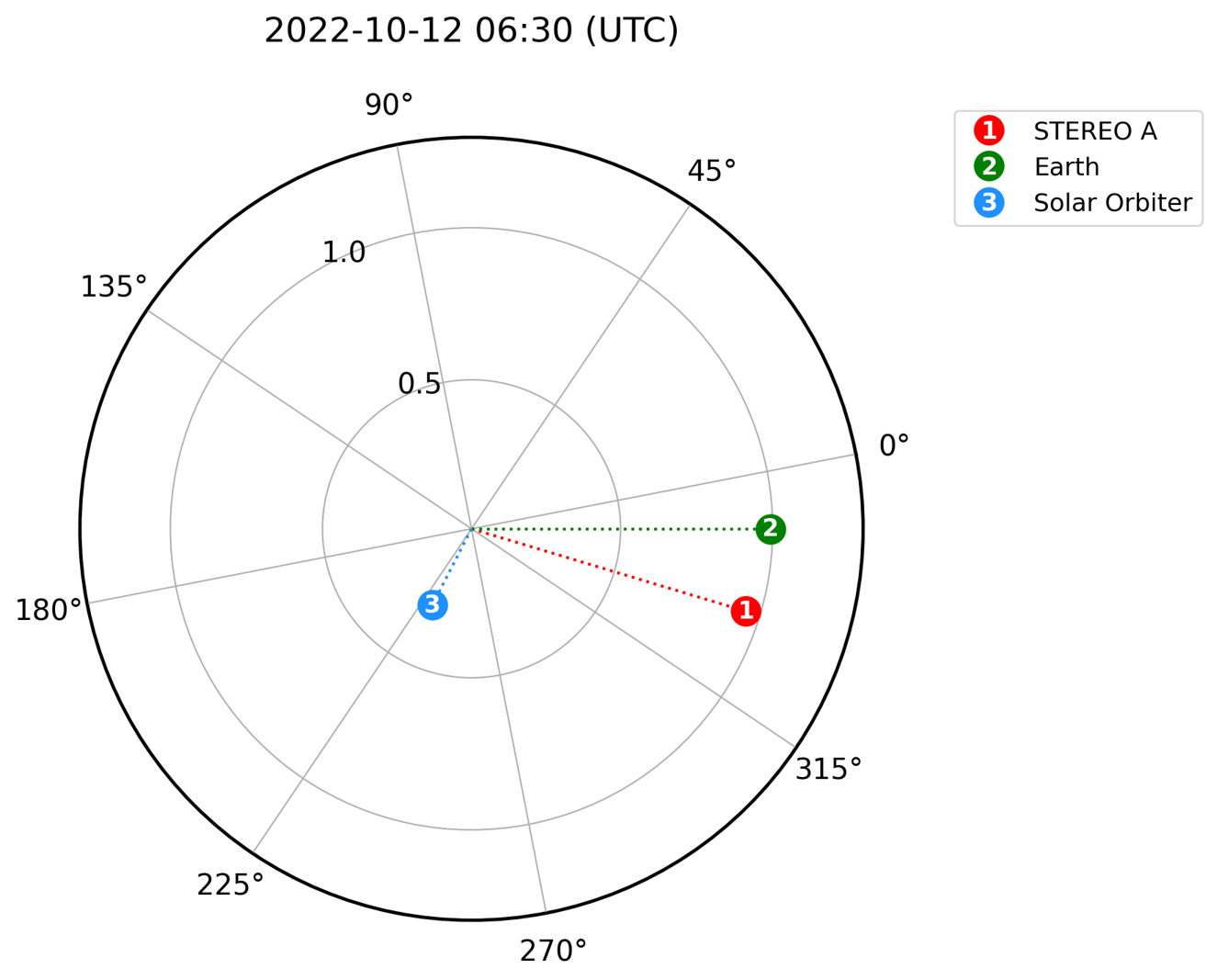}
\caption{Positions of Solar Orbiter and Stereo A with respect to Earth on 12 October 2022, 6:30 UT, in heliocentric Earth equatorial coordinates. The image was created using Solar-MACH (https://solar-mach.github.io/).}
\label{fig1}
\end{center}
\end{figure}

\begin{figure}
\begin{center}
\includegraphics[trim=10 100 50 240, clip, scale=0.42]{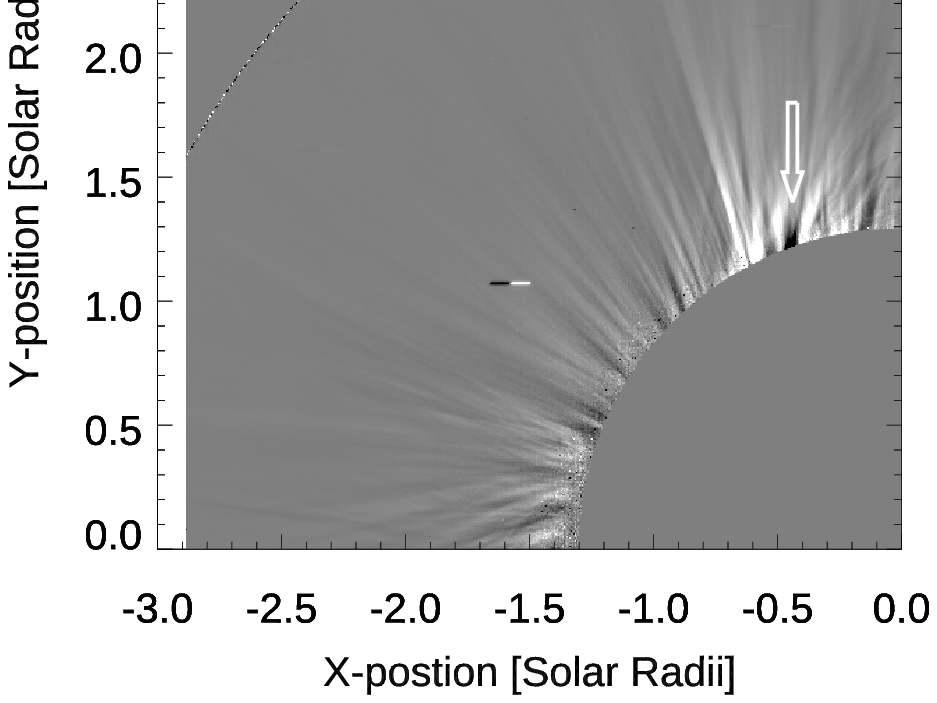}
\includegraphics[trim=10 100 50 240, clip, scale=0.42]{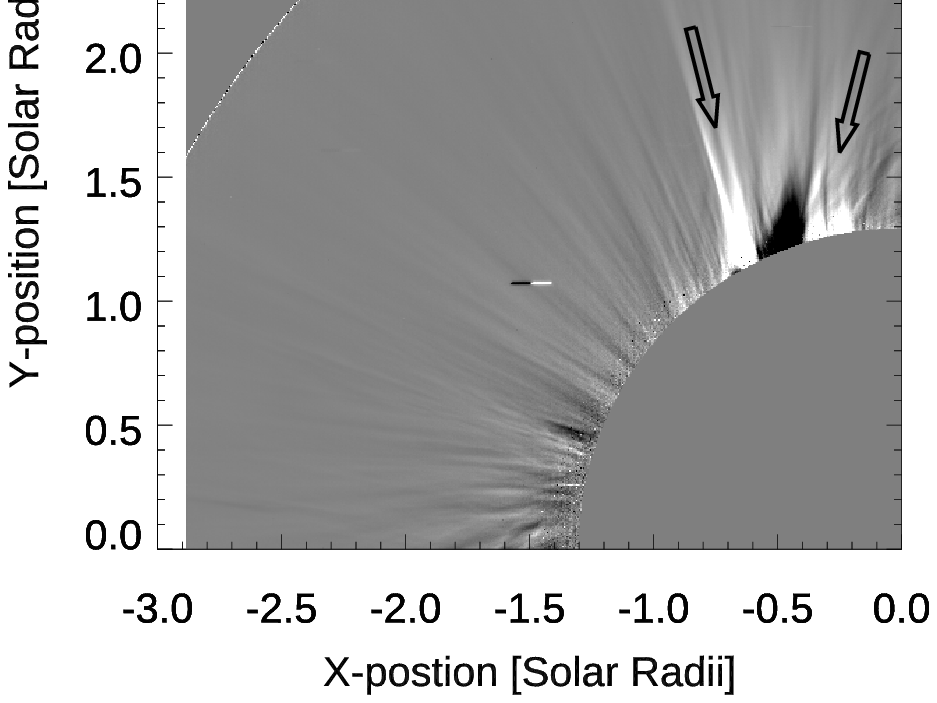}\\
\includegraphics[trim=10 100 50 200, clip, scale=0.42]{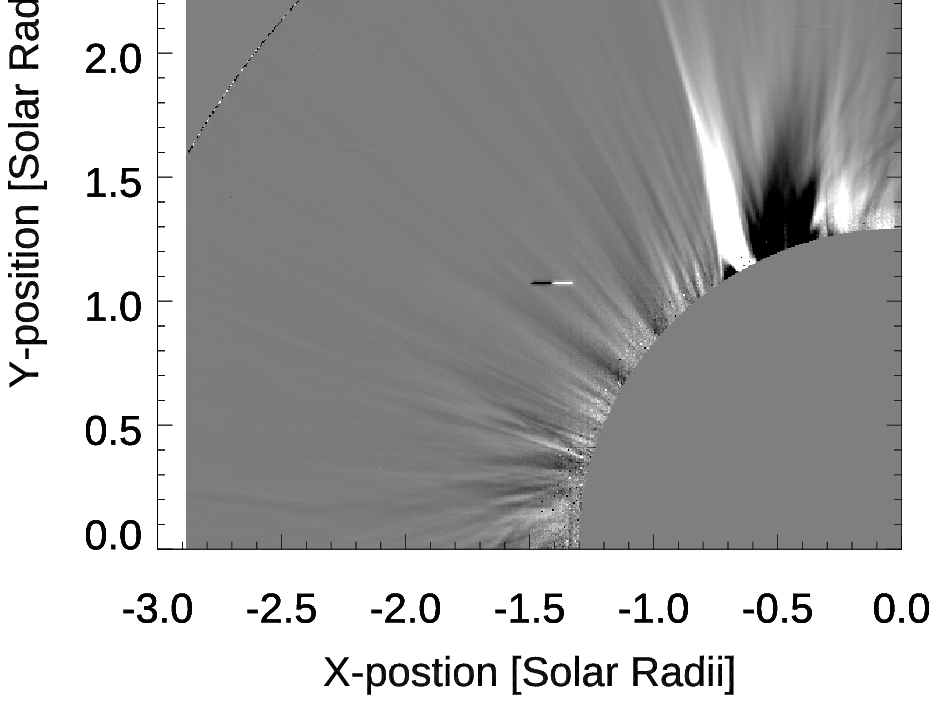}
\includegraphics[trim=10 100 50 200, clip, scale=0.42]{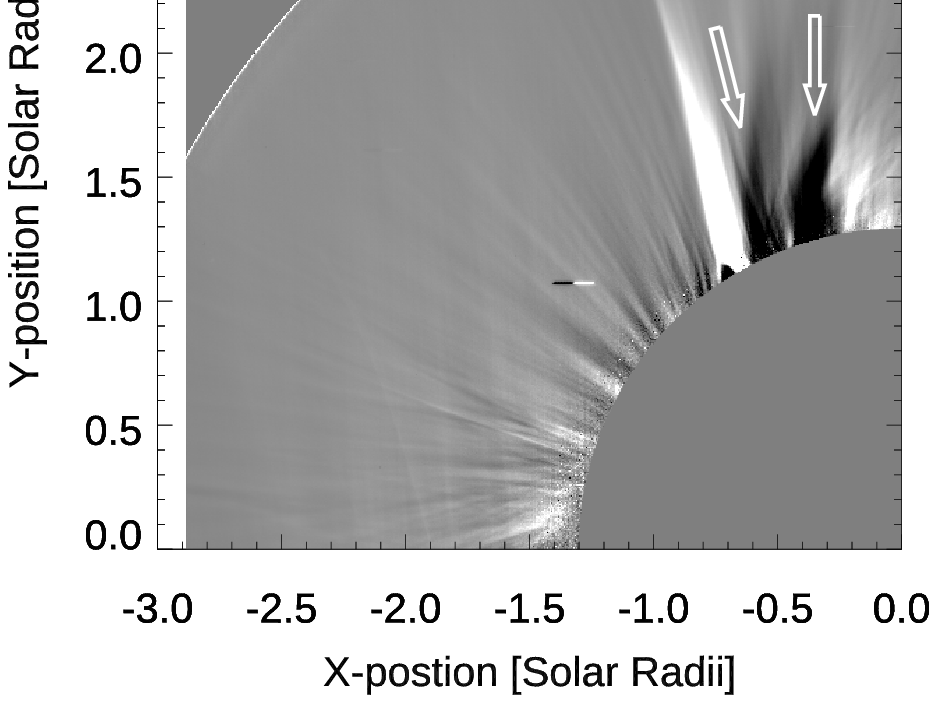}
\caption{Sequence of the normalized running difference maps depicting the propagation of the CME front (white arrows) and the evolution of its two legs (black arrows). The time indicated on each map refers to the image from which the previous one, acquired 16 minutes earlier, has been subtracted.}
\label{fig2}
\end{center}
\end{figure}

\begin{figure}
\begin{center}
\includegraphics[trim=0 120 0 150, clip, scale=0.74]{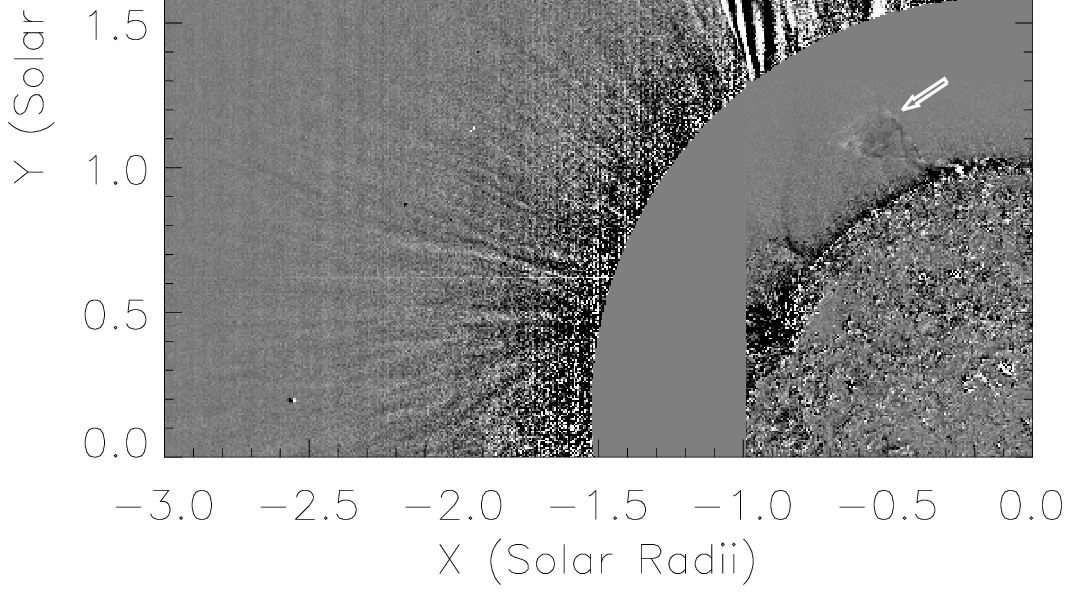}
\caption{Composite image combining an EUI map at 174 \AA{} taken at 6:10 UT and a Metis map captured at 10:17 UT. The normalized running difference technique has been applied to both maps. The arrow points to the eruptive polar crown prominence depicted in Figure \ref{fig3}.
A complete movie covering the EUI observations from October 12 at 04:00 UT to October 13 at 00:30 UT is also available online. The EUI images are processed with Multiscale Gaussian Normalization \citep[MGN,][]{MGN2014}. In the movie a linear feature becomes visible ahead of the erupting prominence at around 5:20 UT. The real-time duration of the movie is 24 s.
A movie showing the persistence of the helical features in the first 3.5 hours of observation of the Metis dataset taken from from 10:15 UT to 20:16 with a time cadence of 2 minutes and 1 second is also available. The real-time duration of the movie is 14 s.
}
\label{fig4}
\end{center}
\end{figure}

\begin{figure}
\begin{center}
\includegraphics[trim=10 180 30 180, clip, scale=0.42]{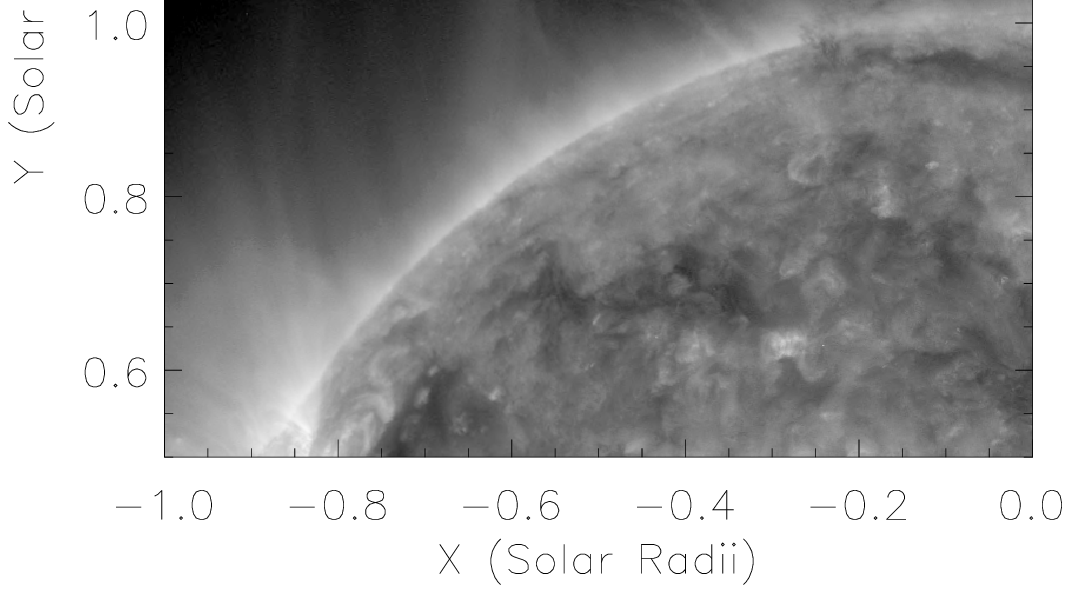}
\includegraphics[trim=40 180 30 180, clip, scale=0.42]{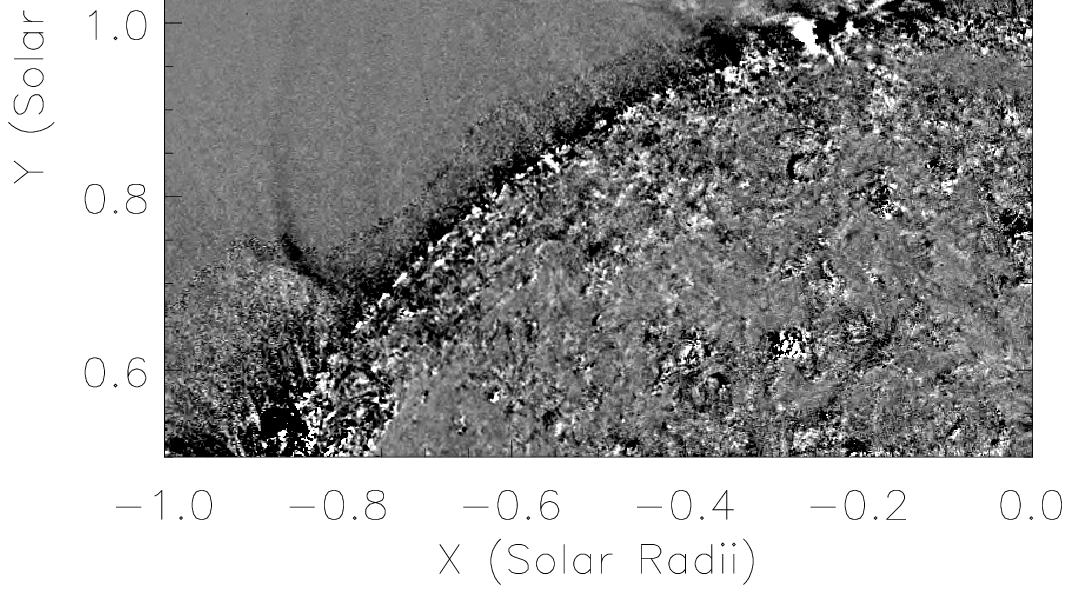}\\
\includegraphics[trim=10 160 30 160, clip, scale=0.42]{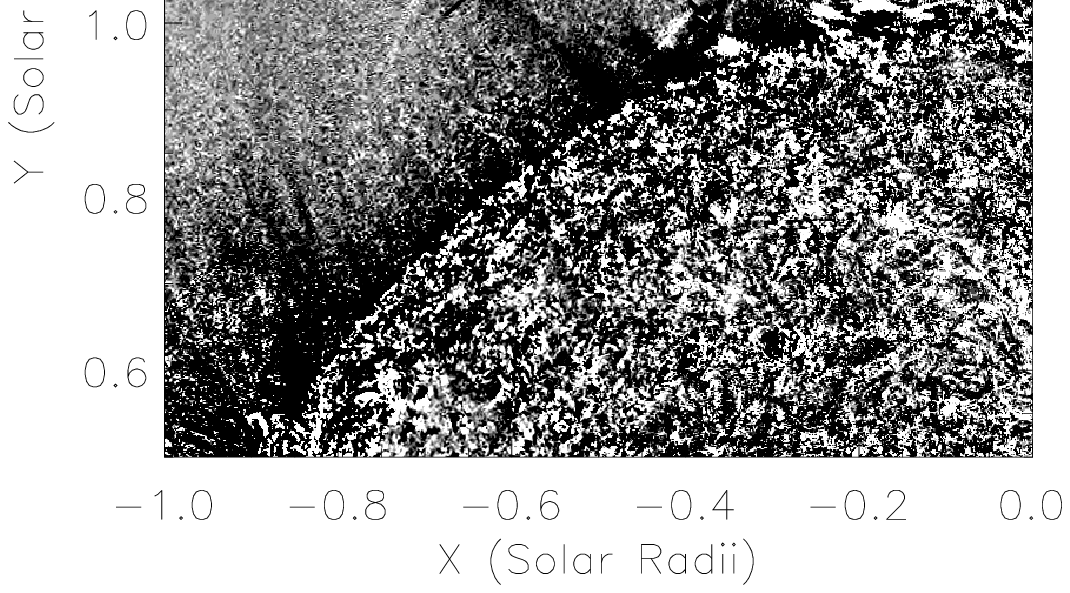}
\includegraphics[trim=40 160 30 160, clip, scale=0.42]{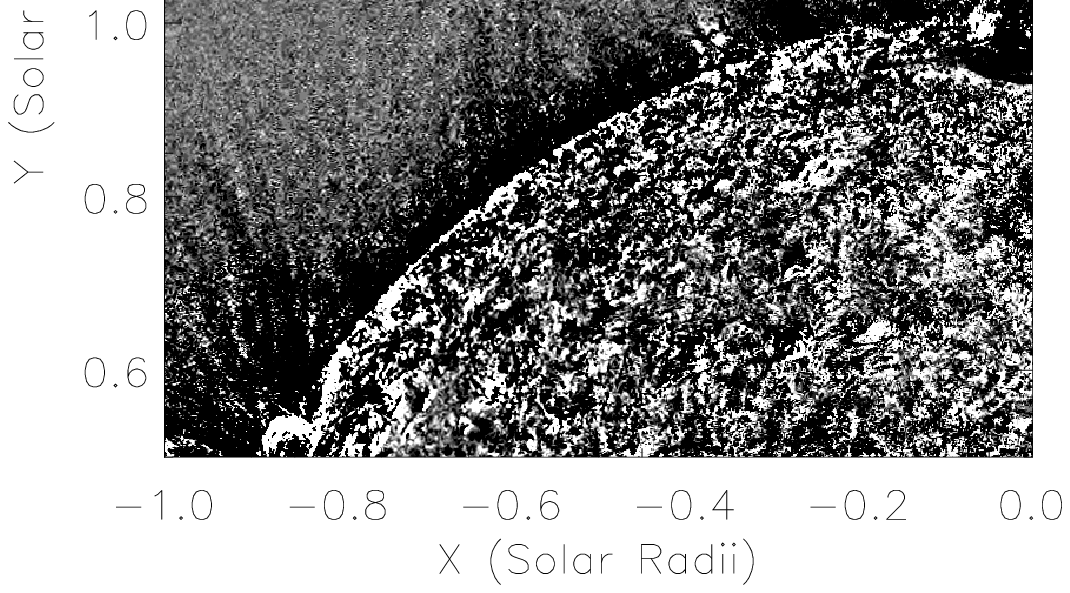}
\caption{In the top-left panel, an EUI map at 174 Å is shown, captured on October 12, 2024, at 6:10 UT. The top-right panel displays the same field of view using a map generated with the running difference technique. Arrows indicate the eruptive polar crown prominence, with its helical configuration clearly visible during the rising phase in the running difference map. The bottom-left and bottom-right panels present running difference maps taken at 6:50 UT and 7:50 UT, respectively.}
\label{fig3}
\end{center}
\end{figure}

\begin{figure}
\begin{center}
\includegraphics[trim=10 30 0 10, clip, rotate=-90, scale=0.5]{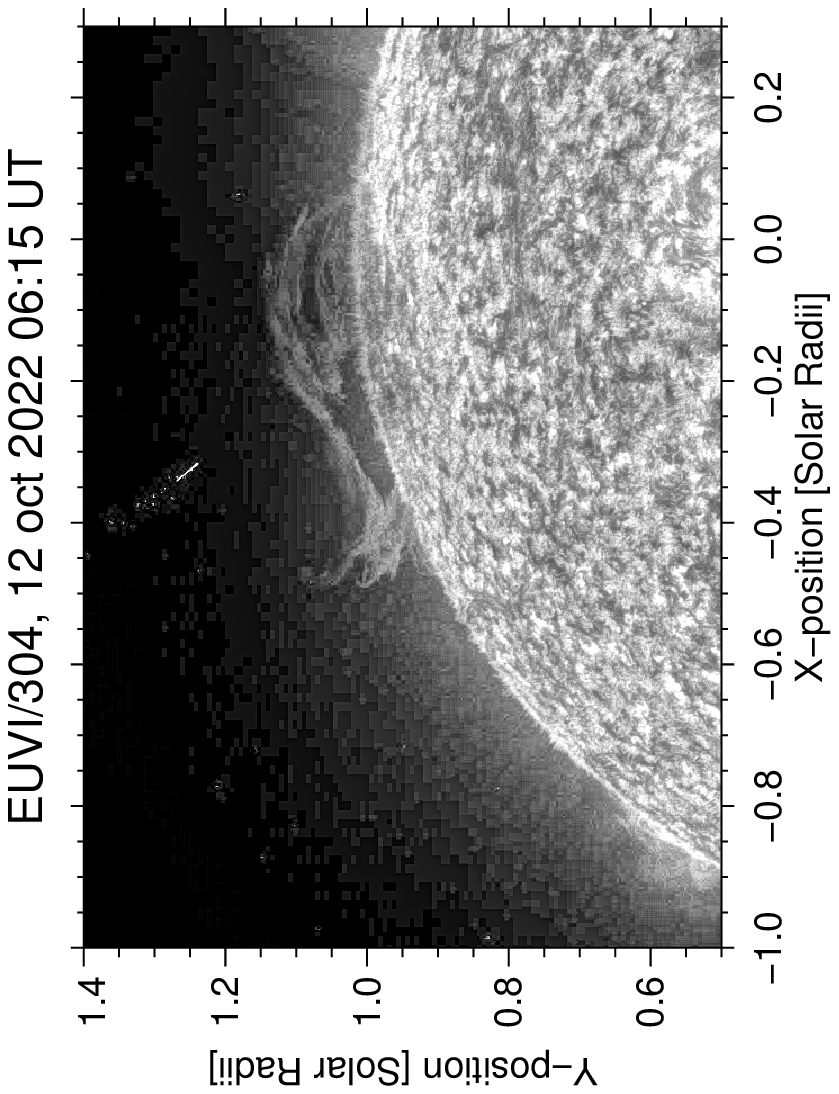}
\caption{STEREO EUVI image at 304 \AA{} of the polar crown prominence taken at 6:15 UT. The image is processed with MGN \citep{MGN2014}.}
\label{fig_EUVI}
\end{center}
\end{figure}

\begin{figure}
\begin{center}
\includegraphics[trim=0 0 0 0, clip, scale=0.5]{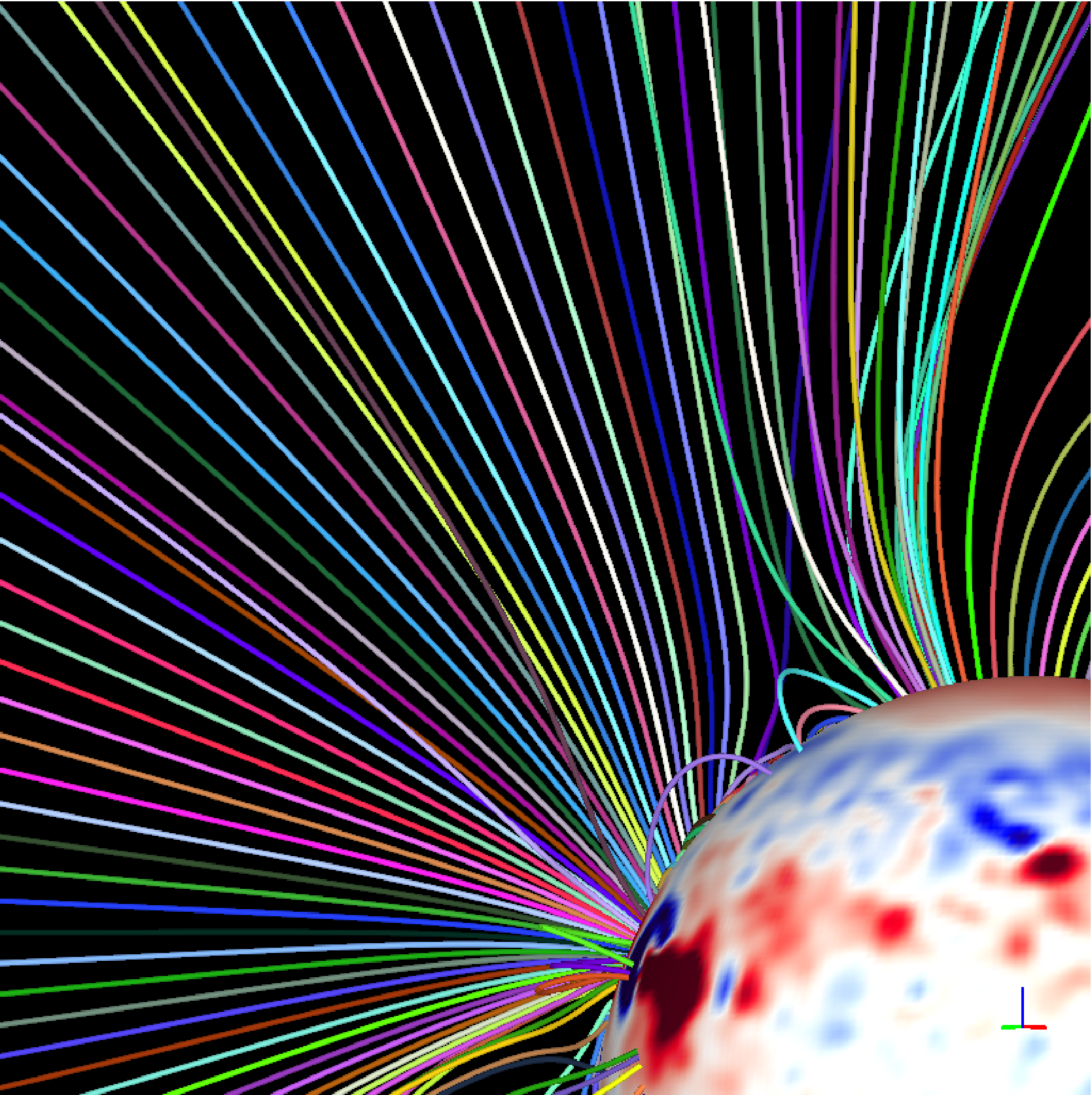}
\caption{Extrapolations of the coronal magnetic field on 12 October 2022. We have identified a pseudo-streamer at the North-East limb at the latitude corresponding to the location of the polar crown prominence. The field of view extends up to 3 solar radii.}
\label{fig5}
\end{center}
\end{figure}

For our analysis, we used Metis VL observations coronagraphic images in visible light acquired on October 12, 2022, from 03:30:01 UT to 23:24:13 UT.  The data were acquired both in polarized-brightness observing mode (VL-pB) and in the so-called ``total brightness'' mode (VL-tB).  The former consists in the acquisition of 4, interlaced polarized VL images that are then combined to obtain a map of the corona in polarized brigthness ($pB$), while the latter acquires the total (unpolarized) coronal brightness through a special hardware mode that switches the polarization angle in the middle of the detector integration time \citep{Ant20}. Table \ref{Tab:data} lists Metis VL observation on that date with their main acquisition parameters.  On that date, Solar Orbiter was at a distance of 0.293~au from the Sun, pointing to a position on the solar surface centered at a Carrington longitude and latitude of about 230~\degr\ and about -3~\degr\ respectively. At that distance, the angular pixel scale of the detector of the Metis VL channel, which is of 10.7\arcsec\ \citep{And21}, corresponding to about 27~Mm on the Sun (54~Mm if a $2\times2$ binning is applied to the data). Figure \ref{fig1} shows the position of Solar Orbiter relative to Earth: as we can see they are almost in quadrature, which means that the Western corona in Metis images lies along the Earth line of sight.  

The highest cadence observation were obtained in the total brightness mode, with cadences of about 20~s from 9:30 to 10:12~UT and about 2 minutes from 10:15 to 20:19~UT.  The data have been calibrated as explained in \cite{Rom21} and \cite{DeL23}.  Further processing steps were applied to the images sequences to emphasize the most rapidly varying features in the field of view.  In particular, for this work we chose to show running difference images normalized by the average radial coronal brightness profiles.  In order to increase the signal-to-noise ratio, we also applied a pixel-by-pixel temporal average to the highest cadence VL-tB acquisition (session no.~228504), effectively reducing the cadence to 1 images every 140~s in that case.

\begin{table*}
  \centering
  \caption{Summary of Metis data sets taken on 2022-10-12 and analyzed in this work: Type of acquisition (``polarized brigthness'', pB, or ``total brightness'', tB), time intervals covered and main parameters of the acquisition session (exposure time, cadence, detector binning).  Each data set is identified by a session number which uniquely identifies in the Metis data base the images of that data set.  Times are UTC onboard.}
  \label{Tab:data}
  \begin{tabular}{r c r r r r r r r}
    \hline\hline
    Index & Type & Session no. & N.\ of images & Start time & End time & Cadence (s) & Binning \\
    \hline
        1 & VL-pB   & 228501      &  19           & 03:30:01   & 08:48:23 & 960         &  2$\times$2 \\
        2 & VL-tB   & 228504      & 120           & 09:30:31   & 10:12:03 &  21         &  2$\times$2 \\
        3 & VL-tB   & 228505      & 299           & 10:15:01   & 20:18:45 & 121         &  2$\times$2 \\
        4 & VL-pB   & 228506      &  36           & 20:25:17   & 23:24:13 & 300         &  1$\times$1 \\
    \hline
  \end{tabular}
\end{table*}

Figure \ref{fig2} reports four $pB$ Metis images processed using the normalized running difference technique between two successive images, acquired with a 16-minute cadence. This sequence reveals the propagation of a jet-like CME at approximately 35 degrees polar angle (measured from Solar North in a counter-clockwise direction). Initially (top left panel of Figure \ref{fig2}), the corona in North-Eastern quadrant appears dominated by radial structures. Starting from 05:38 UT, a dark structure (see the white arrow in top right panel of Figure \ref{fig2}), likely the CME front, becomes visible in the running difference maps, flanked on both sides by two bright features extending to the upper edge of the field of view (FOV) (see the black arrows in top right panel of Figure \ref{fig2}). Note that the easternmost bright radial feature was present before the event, brightens during the eruptions, and as will be shown below, persists well after the CME. We identify this feature as a pseudostreamer  stalk, which corresponds topologically to either an open spine or fan emanating from a coronal null point \citep[e.g.,][]{Wyper2021}. In response to the CME eruption, the bright features seem to gradually move apart (as seen in the bottom panels of Figure \ref{fig2}) with a separation velocity of about 100 km s$^{-1}$ at a distance of approximately 1.5 solar radii. Interestingly, by 06:26 UT, a broad dimming between the bright features has developed (see the white arrows in the bottom right panel of Figure \ref{fig2}). Similar dimmings have been previously reported in jet-like CMEs \citep[e.g.][]{Wang2018,Kumar2021}.

In order to highlight the substructures within the CME and surrounding corona, we applied a normalized base-difference technique. This technique involves computing a base image from the series of images by taking the pixel-by-pixel 1st percentile. The base image is then subtracted from each image, and used to compute a mean radial profile by averaging its values in the azimuthal direction. The image containing the radial profile is then used to normalize the images, enhancing contrast at higher radial distances. Subsequently, we applied the standard running difference technique between two consecutive images \citep{And24}.

Applying the normalised base-difference technique between images with a cadence of 2 minutes and 1 second, at 10:15 UT we observe that the Eastern feature, which we believe corresponds to the pseudostreamer stalk, exhibits a markedly helical configuration along its entire radial extent within the Metis FOV, from about 1.5 to approximately 3 solar radii ($R_{s}$). This configuration persists for more than 3.5 hours from the beginning of the total brightness image sequence (see Figure \ref{fig4}). We also note faint striations in the Western feature and in the gap between them, but the stalk is unique in its pronounced brightness, striking helical dynamics, and persistence. These observations show exactly the type of structure and dynamics that have been postulated by \citet{Raouafi2023} and others, as the origin of the Alfv\'{e}nic solar wind, but on a much larger scale. We discuss this point in more detail directly below.

The observed narrow CME and the continued helical structure appear to be the result of an eruptive polar crown prominence evident in the 174 \AA{} images acquired by EUI \citep{Roc20}, also on board Solar Orbiter, starting at 4:00 UT. These images show the eruption of a prominence at the same latitude near the north solar pole on the eastern limb (top-left panel of Figure \ref{fig3}). Applying the running difference technique to EUI images we highlight the twist of the erupting prominence, manifesting a helical configuration during its rising phase. In particular, at 6:10 UT, the eruptive prominence exhibits an X-shaped structure, with the narrowest part located at a height of approximately 1.1 $R_{s}$ (top-right panel of Figure \ref{fig3}). Ahead of the erupting prominence a linear feature also becomes visible around 5:10 UT and moves upwards and Eastward
disappearing around 7:10 UT (see the online movie). After this time the erupting prominence looses its coherence and fades from view by 8:00 UT. Finally, over the next 12 hours faint cusped loops form in addition to two faint ribbons (bottom-left and right panels of Figure \ref{fig3}).

Taking into account the high latitude from which the CME originates, we might expect that despite the relative positions of Solar Orbiter and STEREO A relative to Earth, the CME and prominence eruption might also be detected. Indeed, the prominence eruption appears above the limb in STEREO EUVI from around 6:10 UT onwards (see Figure \ref{fig_EUVI}), and follows a similar evolution of erupting and loosing its coherence around 7:00 UT. At 7:32 UT a faint jet-like CME appears in the LASCO field of view.

\section{Theoretical interpretation}
\label{sec:theo}

In order to determine the magnetic field configuration of the outer corona in the observed region we used the MAS (Magnetohydrodynamics outside A Sphere) method. MAS is a 3D magnetohydrodynamics (MHD) model \citep{Lin99, Mik99, Mik18} which uses the magnetic flux distribution at the base of the computational domain as boundary conditions, includes thermal conduction along field lines, radiative losses, and coronal heating and integrates the time-dependent equations in spherical coordinates until the configuration reaches a steady state (e.g., \citet{Abb15}, which employed the simpler polytropic approximation). The obtained coronal magnetic field line map extrapolated through the model for the North-Eastern quadrant up to 3 solar radii is shown in Figure \ref{fig5} where we can identify a pseudostreamer at the latitude corresponding to the location of the polar crown prominence. 

The EUI images show that the polar crown prominence resides initially under the northern arch or lobe of this pseudostreamer. The eruption of the prominence follows the typical evolution of a jet-like pseudostreamer eruption, which are themselves very similar to large-scale versions of coronal jets involving small-scale filaments. Both kinds of eruption have been shown to be well explained by the breakout jet model \citep[e.g.][]{Wyper2018}. Here we interpret the linear feature that forms around 5:20 UT ahead of the erupting prominence as a breakout current layer formed at the apparent null point of the pseudostreamer (which is likely a chain of null points and separators, \citep[e.g.][]{Titov2012}). The helical structure within the erupting prominence reflects the expected flux rope structure that could be pre-existing or formed during the eruption with the X-shaped region showing the likely position of the flare reconnection beneath the flux rope. The loss of coherence and disappearance of the erupting prominence is consistent with the reconnection of the prominence flux rope at the breakout current sheet around 7:00 UT. As in previous breakout jet simulation studies \citep[e.g.][]{Wyper2018,Wyper2021} this is expected to transfer some of the twist of the flux rope to the open field, launching helical outflows along the open field consistent with the initial jet-like CME and helical motions observed by Metis.

\begin{figure}
\begin{center}
\includegraphics[width=0.9\textwidth]{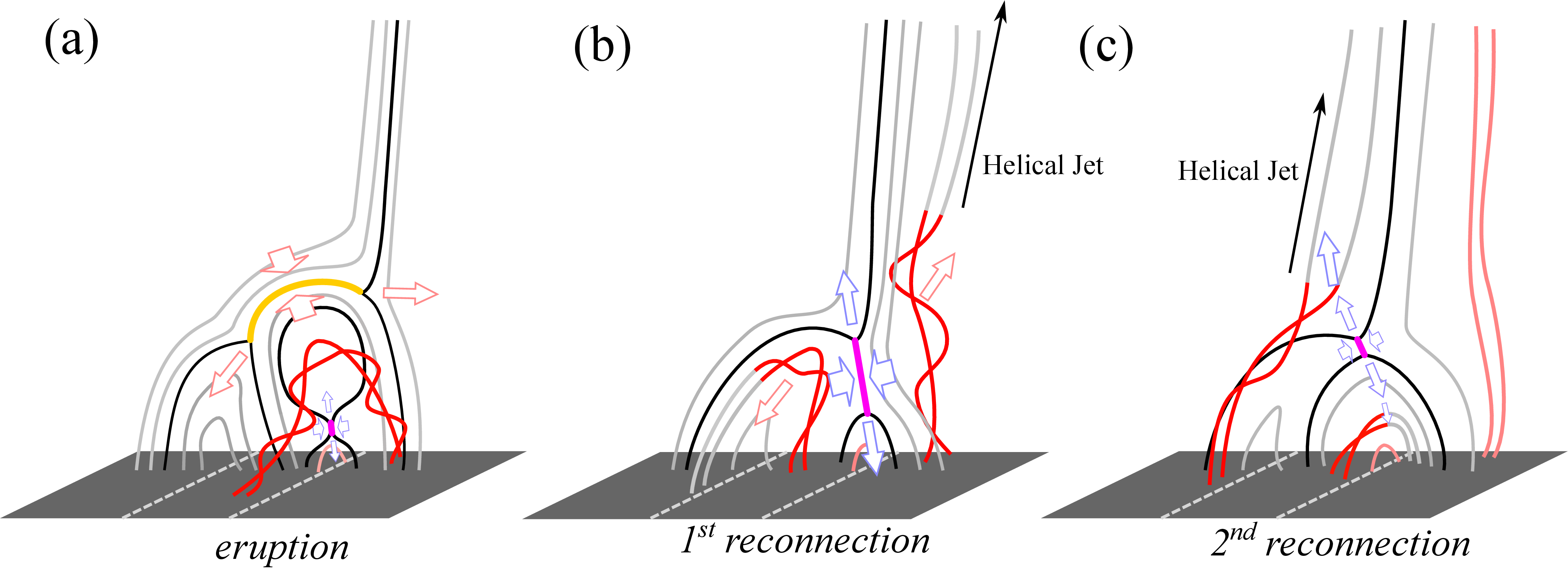}
\caption{Schematic showing a scenario for how helical jet-like motions can arise in the relaxation phase of the breakout jet model. (a) the initial eruption. The breakout and flare current sheets are shown in yellow and magenta, respectively. Arrows indicate the direction of the reconnection and plasma movement in each. Dashed lines show the polarity inversion lines (PILs). (b) the flux rope (red) is split when it reaches the breakout current layer and is reconnected. (c) interchange reconnection in the relaxation phase transfers magnetic flux back over the right PIL, transferring more closed-field twist to the open field line.}
\label{fig:schem}
\end{center}
\end{figure}

\begin{figure}
\begin{center}
\includegraphics[width=0.9\textwidth]{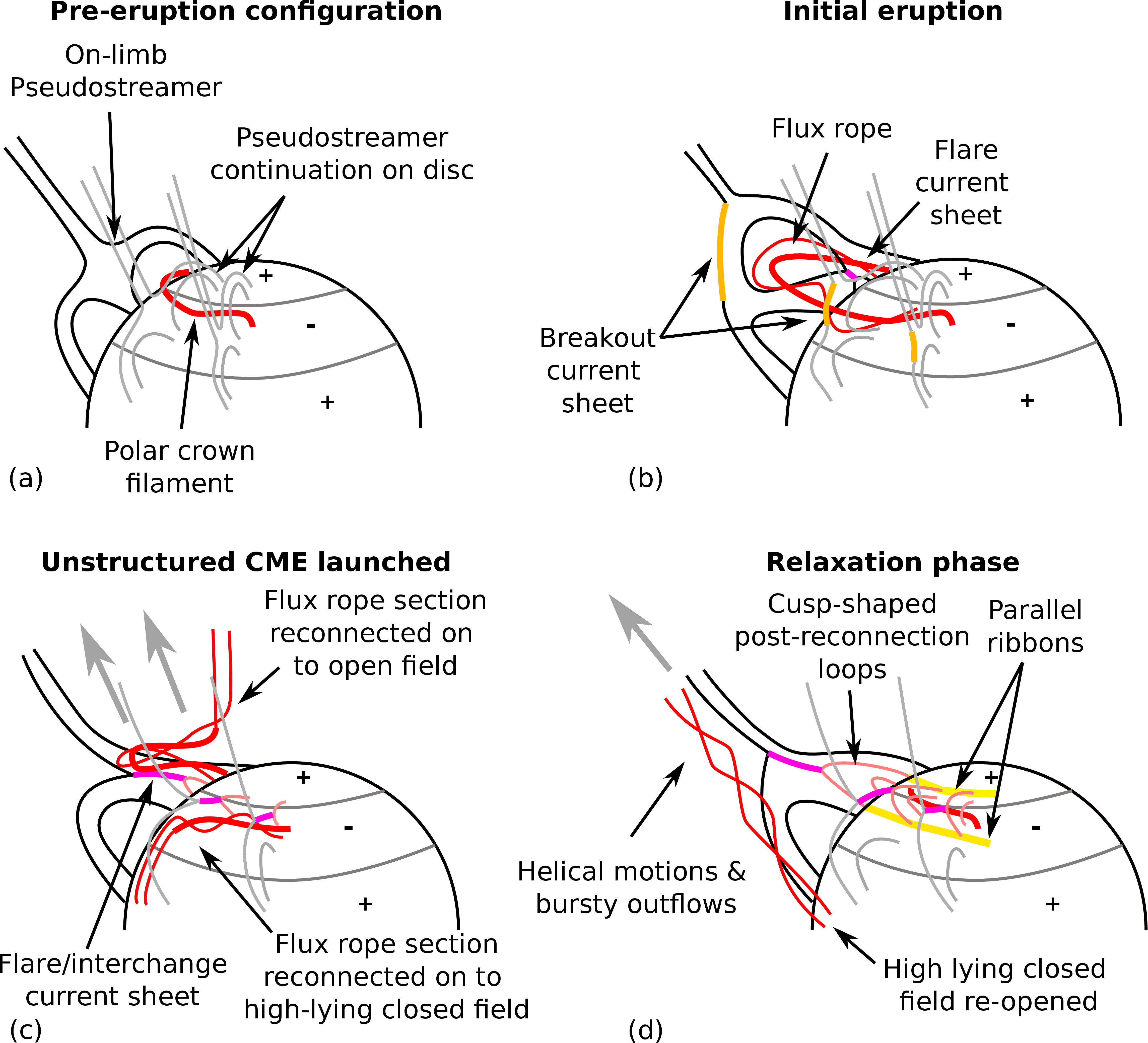}
\caption{Schematic of the hypothesized magnetic field evolution during the event. See \S\ref{sec:theo} for details.}
\label{fig:schem2}
\end{center}
\end{figure}

The observations described above of the erupting prominence and jet-like CME are fairly typical and similar to those of many other events. The striking new features of the Metis observations are the helical motions along the radial stalk that continue to enter the field of view well beyond the initial opening of the flux rope and disappearance of the prominence in EUI around 7:00 UT and outwards propagation of the jet-like CME. These helical motions are still clearly discernible until at least 11:30 UT, far into the relaxation phase of the pseudostreamer after the eruption. We claim that these motions are a result of ``rebound" reconnection of the closed flux system, which is a natural, but until now, unobserved consequence of the evolution expected for a pseudostreamer CME or coronal hole jet. 

We show the basic scenario  schematically in Fig. \ref{fig:schem}. For ease of viewing the system is shown as roughly 2D in that there are two PILs (dashed white lines) on the photosphere and two separate closed arcades. In reality the PILs loop around and form one elliptical PIL, but the schematic captures the basic physics of the system. The breakout and flare reconnection are shown in red and blue respectively in panel (a) during the eruption phase. Panel (b) shows that when the flux rope undergoes reconnection through the null, part of it connects to the open field, which launches the initial jet-like CME, while the rest connects onto the closed field above the left PIL. Note that in this initial reconnection only half or so of the shear/twist in the erupting flux rope escapes out into the heliosphere as a CME, the rest stays inside the pseudostreamer as newly-closed flux overlying the non-erupting portion of the PIL. This result is a basic property of all interchange reconnection; only the stress on the leg of the initially closed field line that becomes open is released into the heliosphere. As a result, the closed field region on the left in Fig. \ref{fig:schem} now has an excess of magnetic flux and shear/twist. Once the filament flux rope exits the pseudostreamer, this closed flux ``rebounds" back toward the right, as in Panel(b), leading to a second round of reconnection, now between the left closed system and open flux on the right. 
This rebound reconnection transfers further twist to the open field, leading to pronounced helical outflows in the relaxation phase. Such double reconnection of the flux rope in breakout simulations has been noted before (see the silver field lines in Fig. 7 of \citet{Wyper2021} for example), but since the simulations were focused on the eruptive phase, this relaxation phase was not explored in detail. Note also that the rebound interchange reconnection again leaves some magnetic shear/twist in the system, so there could be continued back and forth rounds of successively less energetic interchange reconnection. 

We show schematically in Figure \ref{fig:schem2} our interpretation of the event within this framework. The broad region of the CME in the Metis field of view is explained by the jet-like outflows from the eruption following the fan-like open spine of the pseudostreamer. The bright Easternmost feature in Fig. \ref{fig2} corresponds to where the pseudostreamer open spine or fan aligns with the line of sight, while the on-disc extension of the fan-like outflows lead to the relatively broad region of helical and striated outflows (panels (a) to (c)). The helical outflows in the late phase of the event located primarily on the Easternmost edge occur as further twist is released by the rebound reconnection of the closed field above the {\bf southernmost} PIL around the same time that the cusp-shaped loops and faint flare ribbons form. Based on this timeline, we conclude that the Metis observations show outflows and helical motions occurring as a result of long-duration interchange reconnection in the aftermath of the polar crown prominence eruption.

The key point of the rebound or relaxation reconnection is that it is completely generic. The CME, itself, may have unique properties, because it involves flare reconnection and most likely the large deformation of the pseudostreamer topology, but the late-phase evolution corresponds to the interchange reconnection that any parasitic polarity region will inevitably have with surrounding open flux. Even if a bipolar region emerges in a purely potential state, which is highly unlikely, the constant photospheric convective motions are bound to drive interchange reconnection. Both the PSP \citep{Raouafi2023} and very recent Solar Orbiter observations \citep{Chitta2024}, imply that this type of interchange reconnection observed by Metis, but on a much smaller spatial and energy scale, is the origin of the Alfv\'{e}nic solar wind.


\section{Measurements of the helix parameters}

By using total brightness images of the extended corona and applying the running difference technique, we were able to visualize more clearly the helical structure on the easternmost extent of the outflows. 
Specifically, the helical structure emerges from the alternating bright and dark filamentary features that display a greater inclination relative to the longitudinal axis of the flux tube that forms the pseudostreamer stalk.
Here, we define inclination as the angle between these filamentary features and the radial direction originating from the solar disk passing through the midpoint of the intersecting leg of the flux tube and the solar disk. The inclination provides a measure of how much these features deviate from the radial direction. The dark features emerging from the running difference maps were analyzed to determine their inclination.


\begin{figure}
\begin{center}
\includegraphics[trim=40 154 30 280, clip, scale=0.6]{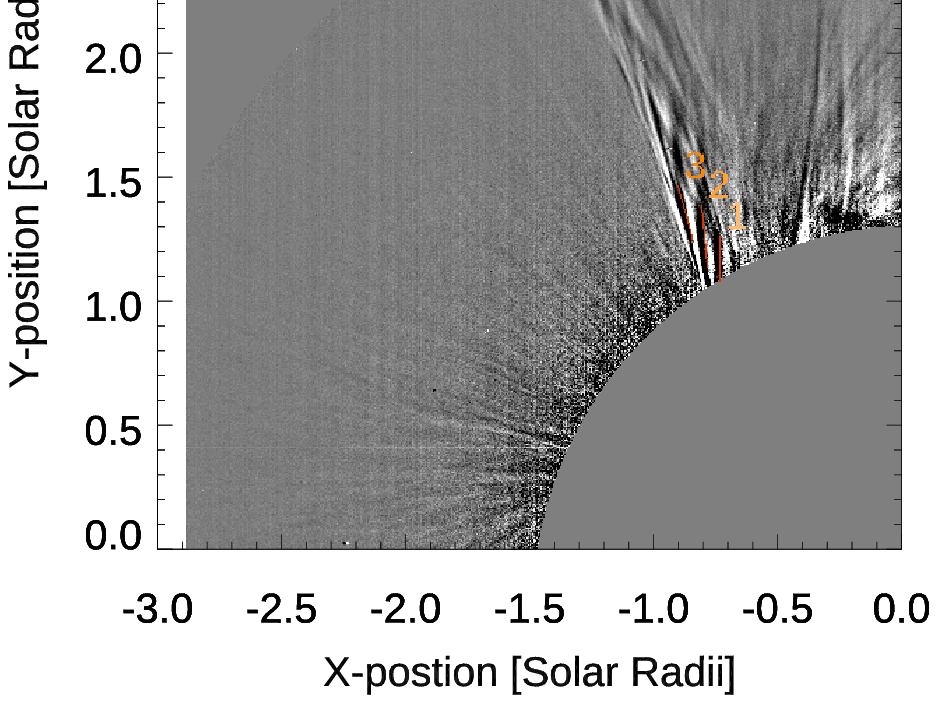}
\includegraphics[trim=40 95 30 280, clip, scale=0.6]{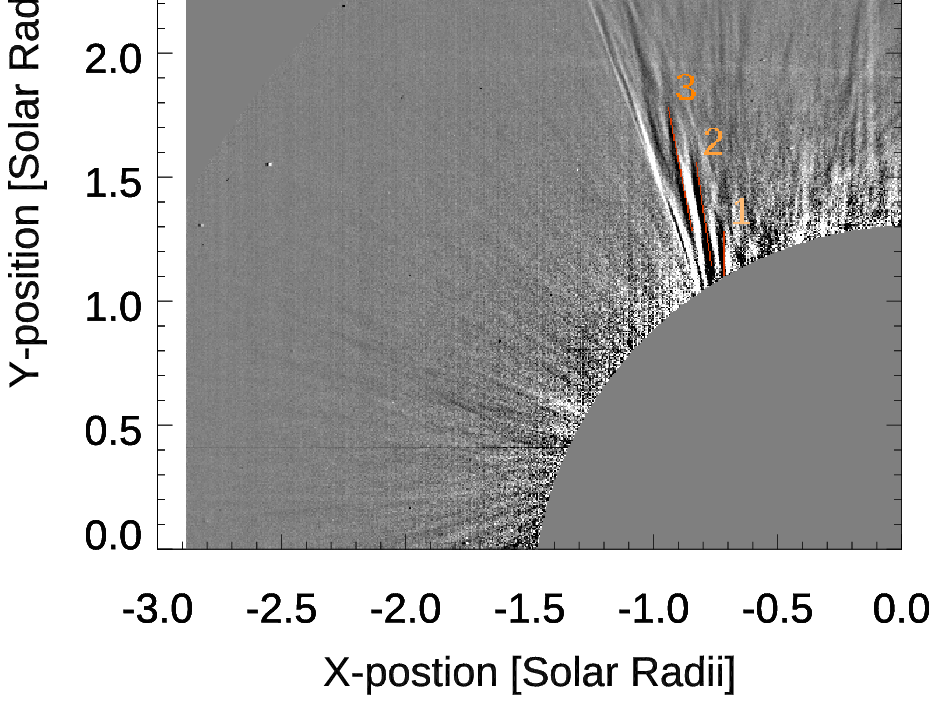}
\caption{Two Metis maps of the north-eastern portion of the corona obtained by the running difference technique. In the top left of each panel, the angle between each structure and the radial direction is indicated. The numbers and the red levels distinguish the different features.}
\label{fig6}
\end{center}
\end{figure}

\begin{figure}
\begin{center}
\includegraphics[trim=40 154 50 400, clip, scale=0.64]{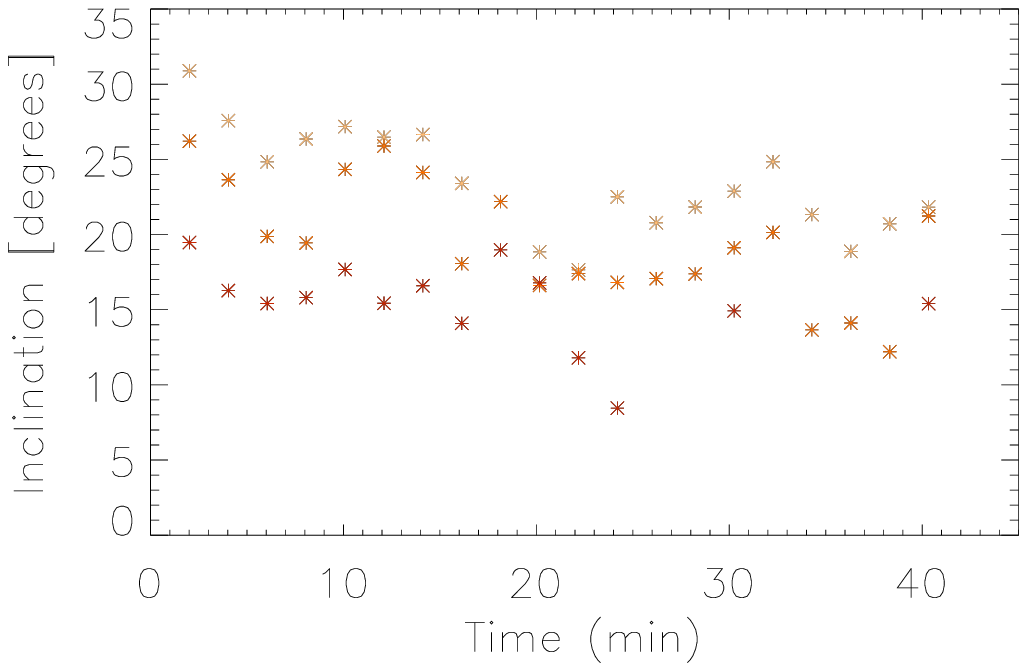}
\includegraphics[trim=40 110 50 380, clip, scale=0.64]{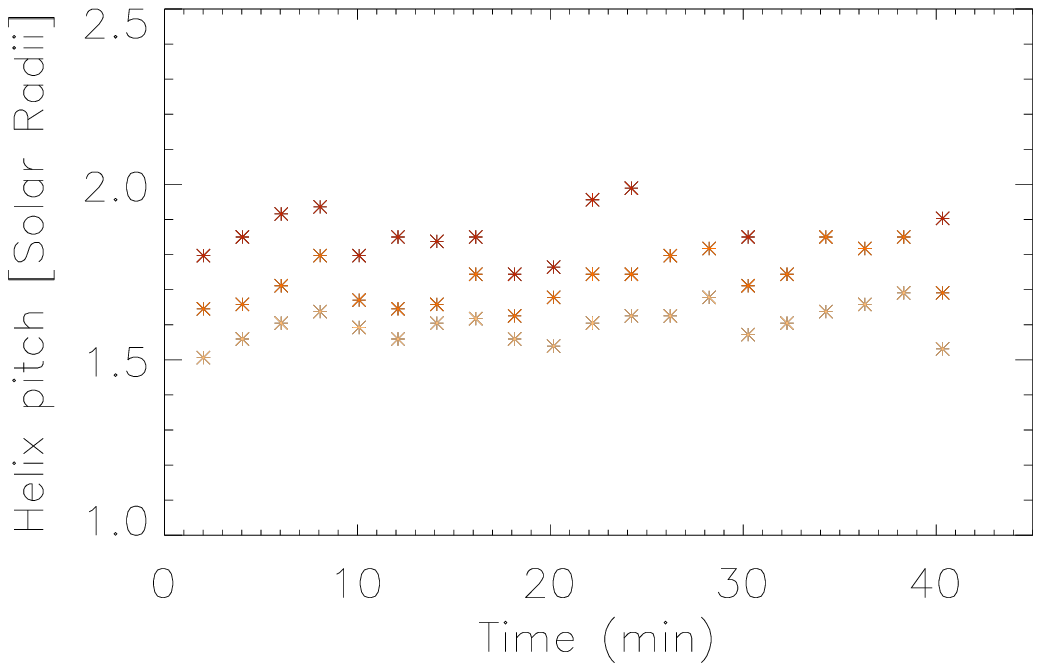}
\caption{Temporal evolution of the inclination and helix pitch of the helicoidal structures, as observed by Metis. t=0 corresponds to 10:15:01 UT on October 12, 2022. The varying shades of red represent features at progressively greater radial distances, with darker shades indicating higher altitudes.}
\label{fig7}
\end{center}
\end{figure}

\begin{figure}
\begin{center}
\includegraphics[trim=40 154 30 380, clip, scale=0.64]{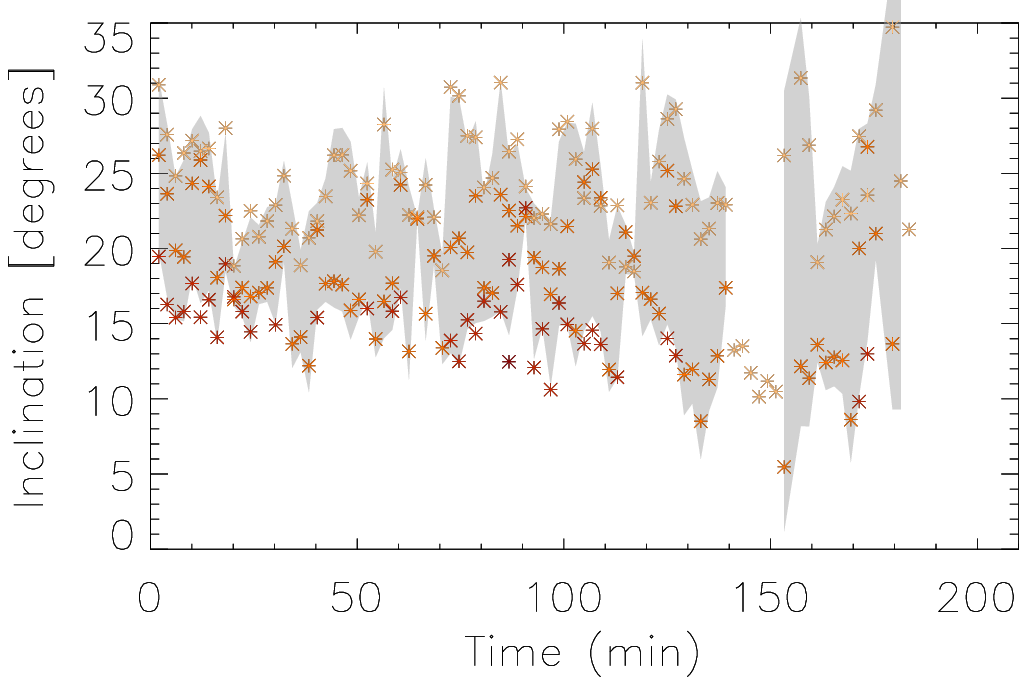}
\includegraphics[trim=40 110 30 380, clip, scale=0.64]{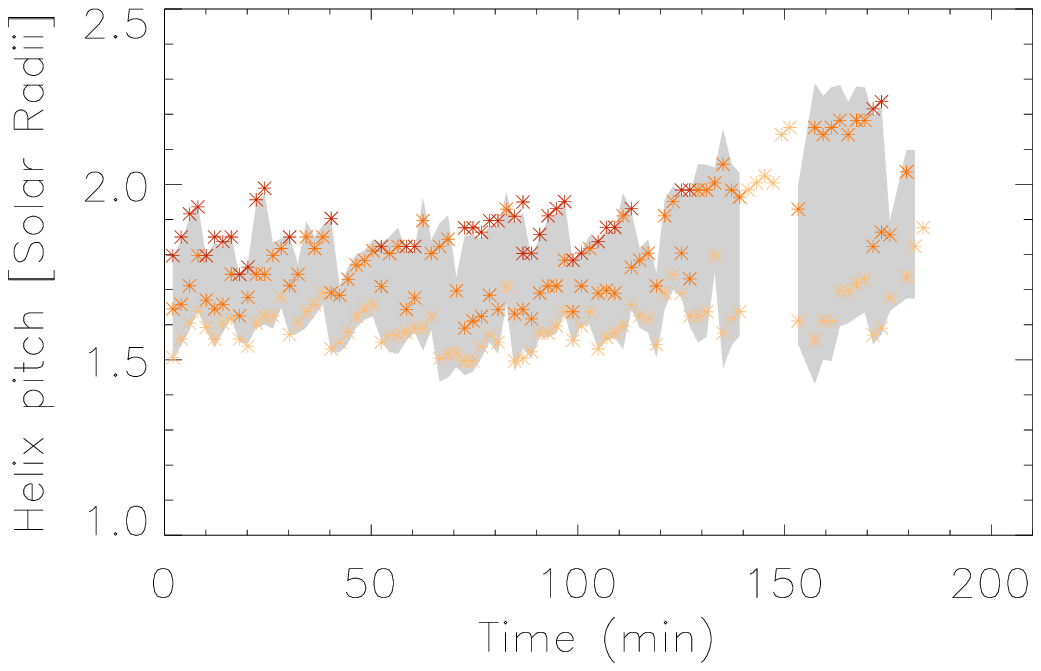}
\caption{Temporal evolution of the inclination and helix pitch of the helicoidal structures, as observed by Metis from 10:15:01 UT (t=0) to 13:36:56 UT on October 12, 2022. The grey regions represent the values corresponding to the average of the quantity $\pm$ the standard deviation.}
\label{fig8}
\end{center}
\end{figure}

In both panels of Figure \ref{fig6}, two running difference maps are shown as examples, with dark features indicated by red segments for which the inclinations are measured and displayed in the top left of each panel. The number of identifiable features varies among frames, and specific features and their evolution from one frame to the next cannot always be consistently tracked. Nonetheless, a consistent trend was observed among the identified features in individual frames, where features located at higher polar angles and higher radial distances exhibit lower inclinations, i.e., closer to the radial direction. In the two frames of Figure \ref{fig6}, the inclination changes from approximately 30 degrees to 16 degrees, a similar trend is visible in the subsequent frames. 

Subsequently, the height of the dark features — referred to as the helix pitch — was measured by determining the distance of the intersection points between these features and the radial direction passing through the midpoint of the intersecting leg of the flux tube and the solar disk. Considering that features at higher coronal polar angles and greater radial distances do not always span a sufficient length to intersect the radial direction defined by the midpoint of the leg, this estimation of the height was performed only for some features in each frame.

The temporal evolution of inclination and helical pitch, utilizing the first 20 available frames, is depicted in the top and bottom panels of Figure \ref{fig7}, respectively. It is found that the average inclination of the helical features varies from 31 to 6 degrees. The span of the radial distance of the intersection points between the dark features and the radial direction remains relatively constant over time, varying between 1.5 and 2.0 $R_{s}$.

Considering a longer time sequence, from the beginning of the available total brightness Metis sequence until the disappearance of the helical structure (around the 100th Metis frame corresponding to 13:22 UT), we note that the inclination of the dark features tends to become more spread out, as highlighted by the standard deviation of the mean shown in grey in the top panel of Figure \ref{fig8}. 
Additionally, there is an increase in the height of the higher intercepts between the dark features and the radial direction, especially towards the end of the sequence, suggestive of an overall reduction in twist within the flux tube towards the end of the observation period (bottom panel of Fig. \ref{fig8}).

Observing the plot in the bottom panel of Fig. \ref{fig8} more closely, we note that some points, likely belonging to the same feature visible in several consecutive frames, display their rising process by continuous monotonic increase of their height: from these sequences of points, we can estimate an ascent velocity of approximately 0.5 $R_{s}$ per hour, corresponding approximately to 100 km s$^{-1}$.

\section{Comparison with an MHD simulation of interchange reconnection}


Considering our interpretation that Metis was observing outflows and helical motions as a result of sustained interchange reconnection, we decided to investigate the similarities between the Metis observations and the results of the high-resolution simulation of interchange reconnection conducted by \citet{Wyp22}. The simulation domain was the volume between between 1 and 20 $R_{s}$ (although the outflows were only resolved as far as 4.5 $R_{s}$), and the system was initialized with a monopolar ambient magnetic field combined with 16 subsurface magnetic dipoles, resulting in a 3D null-point topology similar to the pseudostreamer in this event. The magnetic null-point in the simulation was at a height of about 1.25 $R_{s}$. A fully ionised plasma was assumed with a constant and uniform temperature of 1 MK throughout the computational volume. 

Interchange reconnection was initiated in this simulation by applying twisting motions at the bottom boundary corresponding to the photosphere. Note that the goal of the simulation was to investigate the fundamental process of interchange reconnection, not CME and/or prominence eruption, so the motions were large scale and slow, and simply added a global stress to the closed flux system. As a result of the added magnetic stress, the closed flux expanded upwards deforming the null and separatrix into a current sheet. An important feature of the simulation is that it used adaptive mesh refinement to achieve the highest possible resolution of the current sheet and ensuing reconnection dynamics. Being highly resolved, the current sheet fragmented due to the plasmoid instability and plasmoids with enhanced density repeatedly formed and were ejected from the current layer. Plasmoid formation is expected to be a general property of all high Lundquist number reconnection \citep{Bhattacharjee2009}. As discussed in detail in \citet{Wyp22}, the field line twist within the plasmoids propagated away as torsional Alfv\'{e}n waves. But of most relevance here is that the overdense plasma of the plasmoids then followed behind these waves as field-aligned, dense outflows. Furthermore due to the added magnetic stress the closed field had a shear (non-potential) component. As a result, when this closed field reconnected with the open field the newly formed open field lines had a large-scale kink, i.e., a non-linear Alfv\'{e}n wave, that straightened out over time. Combined with the repeated launching of dense outflows along these field lines by the plasmoid ejections, this led to a curtain of dense filamentary structures that gradually straightened out.



\begin{figure}
\begin{center}
\includegraphics[trim=60 150 60 310, clip, scale=0.64]{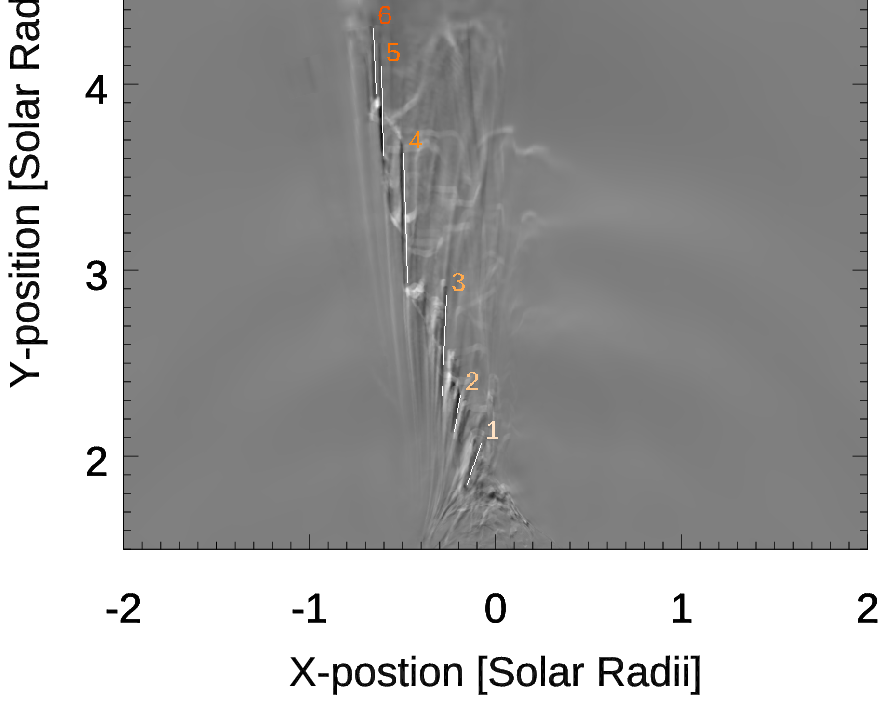}
\includegraphics[trim=60 100 60 310, clip, scale=0.64]{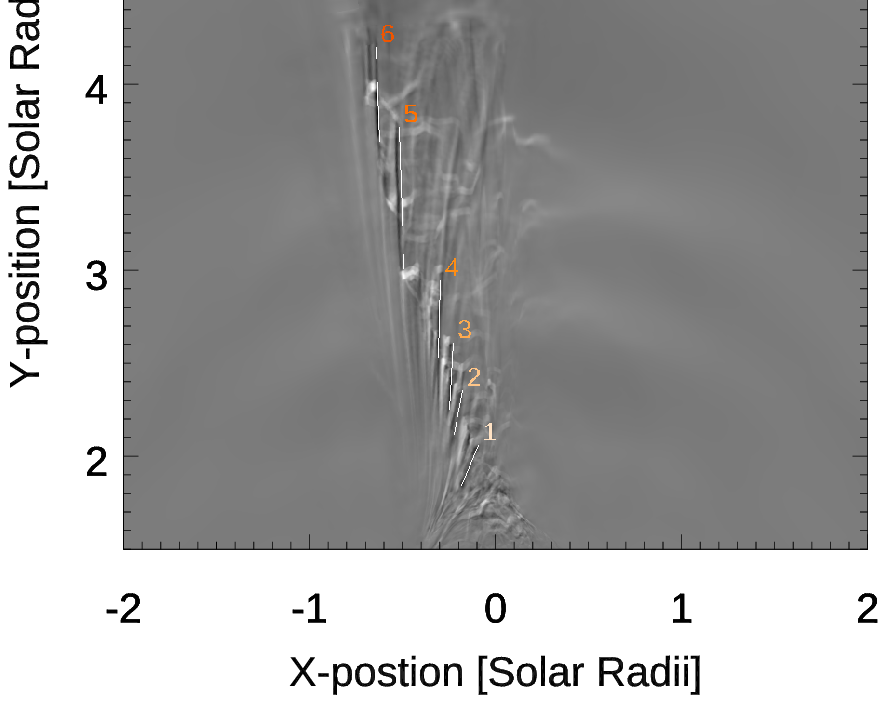}
\caption{Two synthetic white-light images obtained by the \citet{Wyp22} simulation. A measurement of the inclination of the inclined features and their height, similar to the one performed for the observations, was conducted for the simulations. In the top left of each panel, the angle between each structure and the radial direction is indicated.}
\label{fig9}
\end{center}
\end{figure}

When viewed in synthetic white light images derived from the simulation density (see \citet{Lynch2025} for details) these dense features appear very similar to the ones observed by Metis. Therefore, we conducted an analysis similar to the one performed on the Metis dataset, measuring the inclination of the inclined features and the inferred height on the running difference maps derived from the synthetic white-light images created from the simulation. Also in this case, dark features were examined, assuming analogous results for the bright features. As observed in total brightness images, the simulations also exhibit higher inclinations of the lower features located at smaller polar angles. Synthetic images allowed for the segmentation of a greater number of features, taking into account that the outflows are resolved out to 4.5 $R_{}s$ (for example, 5 and 6 dark features have been identified in the frames shown in the top and bottom panels of Figure \ref{fig9}, respectively). In the simulation, the range of inclinations appears broader, spanning from about 28 degrees to -4 degrees (negative inclinations correspond to features forming an angle opposite to the radial direction). However, it is important to acknowledge the limitations inherent in the use of time-differencing techniques to highlight these fine-scale dynamic features. While running difference maps are essential for revealing rapid changes in brightness and detecting propagating structures, they can introduce artifacts or obscure slow-evolving features that may not be clearly visible in individual frames. Additionally, the choice of cadence can influence the visibility and apparent evolution of features, potentially affecting the measured inclination and spatial distribution of the dark and bright structures, although it is noteworthy that the helical features are also visible in the running difference maps obtained using the dataset with a cadence of 21 s (third row of Table 1). Despite these limitations, the consistency observed between the Metis observations and the synthetic images from the MHD simulation reinforces the robustness of our interpretation.

\begin{figure}
\begin{center}
\includegraphics[trim=40 130 30 380, clip, scale=0.64]{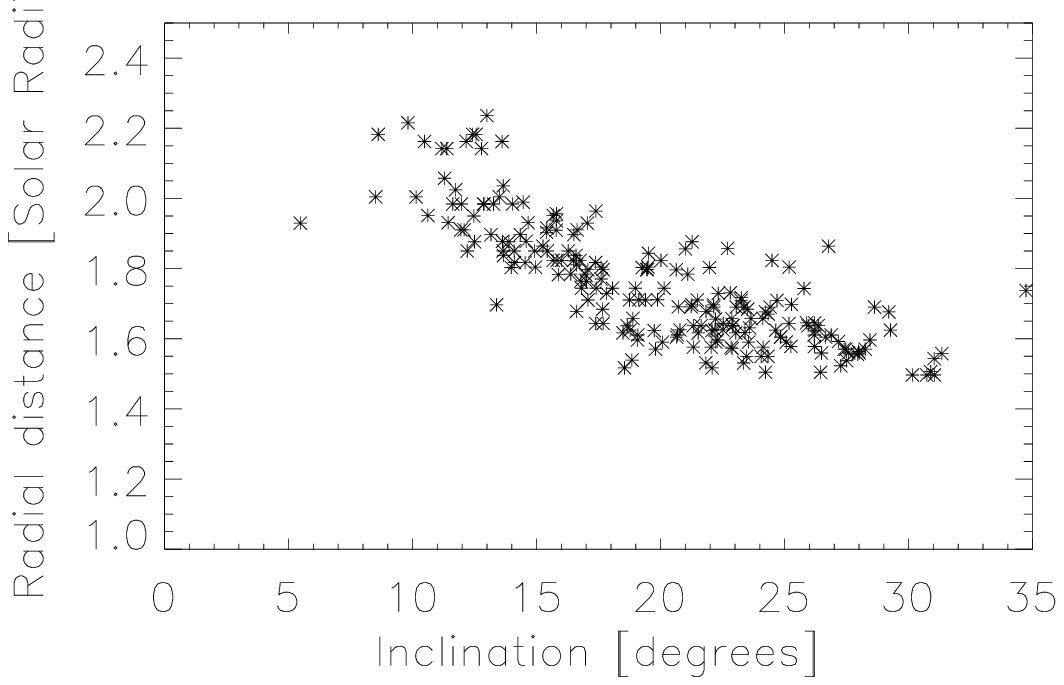}
\includegraphics[trim=40 106 30 380, clip, scale=0.64]{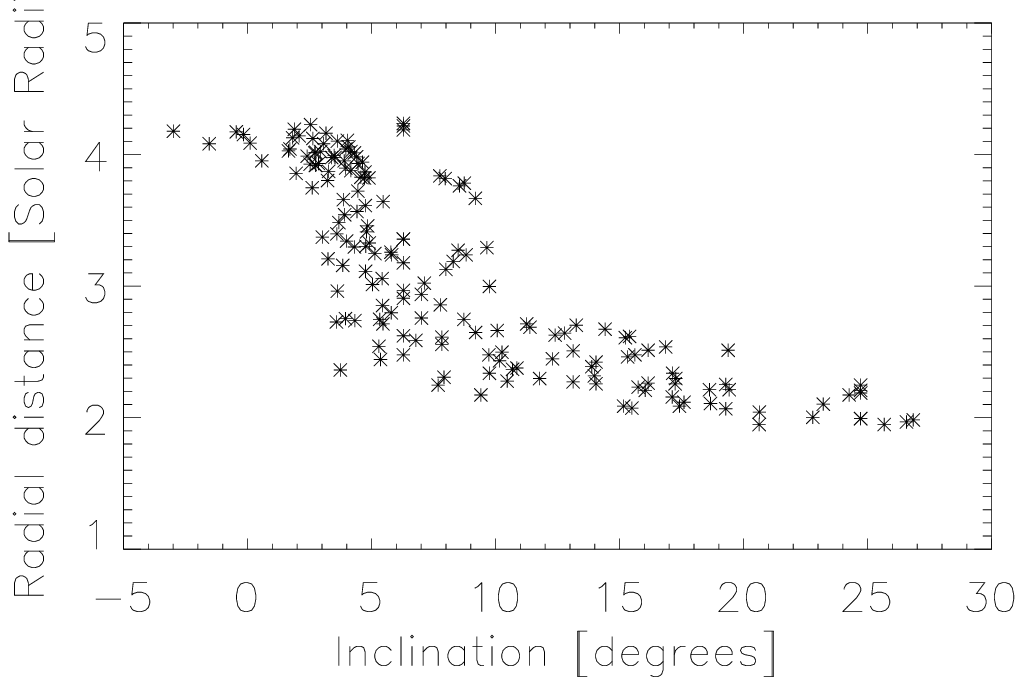}
\caption{Radial distance of the features vs. inclination as determined from the Metis observations (top panel) and the MHD simulation (bottom panel).}
\label{fig10}
\end{center}
\end{figure}

To make a more quantitative comparison between the Metis observations and the simulation, we considered the distribution of the radial distance of the features as a function of their inclination (top and bottom panels of  Figure \ref{fig10}, respectively). In the case of the observations, the radial distance was taken as the intersection of the radial direction with the features, while for the simulation, the midpoint of the features was considered. Although the helical features are located at different altitudes in the observations and the simulation, likely due to the boundary conditions of the latter not being perfectly comparable with the observations, both distributions show a similar anti-correlation between the two quantities, i.e., as the radial distance increases, the inclination decreases. Notably, in the case of the simulation, whose computational domain extends further in terms of radial distance, we observe a slightly more scattered distribution.

Given that the goal of the simulation was a basic investigation of interchange reconnect and was in no way an ``event study", the similarity between the Metis observations and the MHD simulation is remarkable in several aspects. Notably, the persistent inclined features observed across different frames exhibits similar characteristics in terms of their evolution and apparent propagation from one frame to the next, despite the difference in size between the outflowing regions visible in the Metis images and that produced in the simulation. Additionally, we observe that the initial configuration of the MHD simulation resulted in an outflowing region with a substantially radial main axis, similar to what was observed by Metis, but shifted laterally relative to the closed field region due to the twist injected by the driving. In this regard, we remark that the propagation direction of the polar crown prominence observed by EUI manifests a different overall inclination in comparison to the outflows visible in the Metis FOV.

\section{Conclusions}

 In this paper, we present observations by Metis during its perihelion passage of a striking helical radial structure that extended from 1.5 to 3 $R_{s}$ and lasted for more than 3 hours. Thanks to the high resolution of the images and a long-duration dataset (about 10 hours) with a time sampling of 2 minutes and 1 seconds, we were able to measure the evolution of the helical structure and its fine-scale features. To our knowledge, these observations are unique in that they appear to show directly the long-duration outflow of Alfv\'{e}nic solar wind into the heliosphere. The observed features forming the helical structure are likely the result of bursty dense outflows and closed-field twist launched by interchange reconnection occurring in the wake of a jet-like CME, as indicated by the corresponding EUI images. An important finding is the long duration of the helical outflows, which persist until well after the initial eruption shown in EUI. We have shown that this can be explained naturally by the relaxation process expected within the framework of the breakout jet model in the topology of a large pseudostreamer \citep{Wyper2018}.

The spatial and temporal variation of the helical pitch and the inclination of the visible features allow us to interpret the helical structure as an open magnetic flux tube characterized by an untwisting configuration of the magnetic field and corresponding to the stalk of the pseudostreamer. The fact that features located at higher polar angles and higher radial distances exhibit lower inclinations suggests that the field lines along which these features flow are becoming more radially aligned with distance and/or exhibit rotational motion across the plane of sky and then into the plane.
Additionally, the presence of bright and dark features northward away from the easternmost helical features suggests a large portion of the pseudostreamer is involved with a fan-like outflow following its open fan-like spine.


The comparison between the Metis observations and the MHD simulations of \citet{Wyp22} reveals striking similarities, particularly in the geometric parameters and evolution of these helical features. The magnetic field extrapolation, which detected a pseudostreamer at the same location as the helical structure and in agreement with the magnetic configuration of the simulation, reinforces the validity of this comparison. These similarities suggest that the observed features and their persistence for several hours are the result of long duration interchange reconnection, as is consistent with the simulation. Furthermore, the characteristic black and white stripes in the synthetic white-light running difference images derived from the simulation demonstrate a new and straightforward to identify signature of bursty interchange reconnection which due to the high time cadence of Metis appear to have been observationally verified we believe for the first time.


Furthermore, considering that the \citet{Wyp22} simulation provides a potential explanation for the origin of waves and fast outflows that evolve to produce the observed strong drops or reversals of the radial magnetic field component in the solar wind, accompanied by spikes in radial velocity \citep[magnetic switchbacks,][]{Bal19, Kas19}, we expect that events similar to the one reported in this work are critical for understanding the formation, evolution, and eventual dissipation of localized magnetic structures in the solar wind.
One possibility is that the reconnection processes underlying our unique observations could serve as sources of perturbations that, as they propagate, become amplified or give rise to instabilities within the solar wind. Our observations, therefore, might be revealing the origins of the switchbacks/microstreams observed by the Parker Solar Probe mission, but many more studies are needed to clarify these connections.

In conclusion, the observations made by Metis during its perihelion passage provide unique and important insights into the basic dynamics shaping the outer corona and inner heliosphere. The consistency between these observations and the MHD simulations demonstrates that interchange reconnection is the crucial process for the formation of helical structures that are signatures of torsional Alfv\'{e}n waves and solar wind outflows. We emphasize that our observations and simulation are generic in that they capture the fundamental process of interchange reconnection associated with the release of magnetic stress in the closed field region of an embedded bipole. The only significant physical difference between our jet CME event and the ubiquitous jets and microjets that have been proposed as the source of the Alfv\'{e}nic wind is spatial scale. Our results imply that long-lived Alfv\'{e}nic outflows should be a universal feature of the smaller jets, as well. On the other hand, spatial scale is likely to be highly significant in determining whether the outflows contribute to the wind or simply fall back down to the chromosphere. The outflows that we observe start high up and appear to escape the Sun, but this may not be the case for the outflows from very-low-lying jets in the network. We conclude, therefore, that a definitive understanding of the generation of the Alfv\'{e}nic solar wind requires many further studies. This conclusion underscores the importance of future high-resolution, long-duration coronal observations for advancing our knowledge of solar wind origin and evolution.

\acknowledgments
The Metis programme is supported by the Italian Space Agency (ASI) under the
contracts to the co-financing National Institute of Astrophysics (INAF): Accordi
ASI-INAF N. I-043-10-0 and Addendum N. I-013-12-0/1, Accordo ASI-INAF
N.2018-30-HH.0, and under the contracts to the industrial partners OHB Italia
SpA, Thales Alenia Space Italia SpA and ALTEC: ASI-TASI N. I-037-11-0 and
ASI-ATI N. 2013-057-I.0.
The Metis team thanks the former PI, Ester Antonucci, for leading the development of Metis until the final delivery to ESA. P.R. also wishes to thank her for the helpful comments on the paper.

Solar Orbiter is a space mission of international collaboration between ESA and NASA, operated by ESA. Metis was built and operated with funding from the Italian Space Agency (ASI), under contracts to the National Institute of Astrophysics (INAF) and industrial partners. Metis was built with hardware contributions from Germany (Bundesministerium für Wirtschaft und Energie through DLR), from the Czech Republic (PRODEX) and from ESA.
 
This work was supported by INAF (Bando per il finanziamento della Ricerca Fondamentale 2022 - Study of the correlation between the solar activity and the geomagnetically induced currents in gas pipelines systems - and Bando per il finanziamento della Ricerca Fondamentale 2023 - IDEASW project), by ASI under contract with INAF no. 2021-12-HH.0 “Missione Solar-C EUVST – Supporto scientifico di Fase B/C/D” and no. 2022-29-HH.0 "MUSE".
 
PFW acknowledges support from STFC (UK) consortium grant ST/W00108X/1 and a Leverhulme Trust Research Project grant.
SKA acknowledges support from the NASA LWS and the NSF SHINE Programs.
We acknowledge the use of the different facilities, databases and tools appearing in our paper: SOLO (EUI, Metis).\\









\clearpage


\end{document}